# Quantum prime factorization algorithms using binary carry propagation

**Arim Ryou[1], Kiwoong Kim[1], and Kyungtaek Jun[2,3]**

[1]Department of Physics, Chungbuk National University, Chungcheongbuk-do, 28535, South Korea
[2]Quantum Research Center, QTomo, Chungcheongbuk-do, 28535, South Korea
[3]Chungbuk Quantum Research Center, Chungbuk National University, Chungcheongbuk-do, 28644, South Korea

Corresponding author: Kiwoong Kim (email: kiwoong@chungbuk.ac.kr) and Kyungtaek Jun (email: ktfriends@gmail.com).

## Abstract

The RSA cryptosystem, which relies on the computational difficulty of prime factorization, faces growing challenges with the advancement of quantum computing. In this study, we propose a quantum annealing-based approach to integer factorization using both high-order unconstrained binary optimization (HUBO) and constrained quadratic model (CQM) formulations. We begin by modeling binary multiplication with explicit carry propagation, translating this into a HUBO representation and subsequently reducing it to a quadratic unconstrained binary optimization form compatible with current quantum solvers. To address scalability limitations, we implement a CQM approach with constraint relaxation and global product consistency. While the HUBO model successfully factors small semiprimes, it exhibits exponential memory growth, making it impractical for inputs larger than 10 bits. In contrast, the CQM model achieves accurate factorization of semiprimes up to 60 bits—including N = 1,152,921,423,002,469,787—demonstrating significantly improved scalability. Experimental results further show that applying global product constraints enhances factorization accuracy and consistency across all tested instances. This work highlights both the promise and current limitations of quantum-assisted factorization and establishes a foundation for evaluating RSA security in the emerging quantum era.

## Introduction

The RSA cryptosystem is a foundational public-key cryptographic algorithm designed for secure data transmission in digital communications [1]. It relies on the mathematical principle that, given a sufficiently large semiprime number $N = p \times q$, where $p$ and $q$ are distinct large prime numbers, it is computationally infeasible to determine the original factors from the product. The RSA algorithm utilizes a key pair: a publicly shared key used for encryption and a private key known only to the receiver for decryption. The security of RSA is fundamentally anchored in the difficulty of prime factorization—specifically, the challenge of decomposing a large semiprime $N$ into its prime constituents. Although conceptually straightforward, the factorization problem becomes extremely complex as the size of $N$ increases. The most efficient classical approach to date, the General Number Field Sieve (GNFS), exhibits sub-exponential but still super-polynomial runtime complexity in the bit-length of $N$, making it practically unfeasible for large integers [2]. Specifically, the GNFS requires a runtime of $\exp\left((\log N)^{\frac{1}{3}}(\log \log N)^{\frac{2}{3}}\right)$, which scales prohibitively with growing key sizes. Due to this computational intractability, large semiprime factorization is widely believed to be within the class of NP-complete or NP-intermediate problems—problems for which no known polynomial-time algorithms exist on classical computers. This assumption underpins the reliability of RSA in ensuring data confidentiality, authentication, and integrity across various domains, including e-commerce, secure communications, and digital signatures [3]. The robustness of RSA has made it a cornerstone of modern cryptographic infrastructure, despite the increasing pressure to adapt in the face of emerging quantum threats.

With the rapid development of quantum computing, however, this foundational assumption is being challenged. Quantum computers operate based on superposition, entanglement, and interference, enabling fundamentally different computational strategies. Shor's algorithm, in particular, demonstrated that integer factorization can be performed in polynomial time on a fault-tolerant quantum computer [4], thereby threatening the core security assumption of RSA. While experimental demonstrations of Shor's algorithm have been performed for small integers [5], its application to large semiprimes remains constrained due to limitations in current universal quantum hardware. In parallel, alternative quantum paradigms such as quantum annealing have gained traction. The D-Wave quantum annealer, which solves quadratic unconstrained binary optimization (QUBO) or Ising models [6], presents a near-term route to explore quantum-enhanced factorization. Dridi et al. demonstrated factoring semiprimes up to 200,099 using Gröbner basis methods on a D-Wave 2X system [7], though these approaches face scalability issues due to the exponential growth of symbolic computation. Jiang et al. extended this work by embedding logical qubits with full connectivity, factoring numbers up to 376,289 on the D-Wave 2000Q system using 94 logical qubits [8]. With the introduction of hybrid solvers capable of handling up to a million variables [9], more scalable approaches have become viable. Recent studies, such as those by Jun and Lee [10,11], applied higher-order unconstrained binary optimization (HUBO) models to represent $|N - pq|^2$, encoding prime factors $p$ and $q$ in binary (radix-2) form [12]. While the HUBO approach provides high logical fidelity, it suffers from coefficient scaling issues when translated into QUBO form—particularly due to hardware constraints on the precision of coefficient values. These limitations hinder large-number factorization, thus maintaining the practical security of RSA even in the quantum era [10].

In this paper, we propose a binary-constrained formulation of prime factorization that leverages both HUBO and constrained quadratic model (CQM) approaches [13]. We begin by formulating the HUBO model using radix-2 binary multiplication with explicit modeling of carry propagation. To address scalability issues, we transform the HUBO model into a QUBO structure using hybrid solvers capable of handling more qubit variables. Finally, we apply a CQM-based method using constraint relaxation and global product consistency, enabling the successful factorization of semiprimes up to N= 1,152,921,423,002,469,787 (= 1,073,741,789 × 1,073,741,783). Our results demonstrate that while RSA remains secure under current quantum limitations, the CQM formulation offers a scalable and robust path toward quantum-assisted factorization.

## Method

### Binary Multiplication and Carry Propagation

Prime factorization of a large semiprime number $N$, where $N = p \times q$, requires a detailed understanding of the structural relationship between the prime multiplicands $p$ and $q$, and their product $N$. A particularly insightful approach involves representing this multiplication process in binary form, allowing for a bitwise analysis that explicitly considers carry propagation. This binary perspective forms the foundation for subsequent modeling techniques, including constraint-based formulation and factor inference methods.

Let the radix 2 representation of $p$ and $q$ be given by:

$$p = \sum_{j=0}^{n} p_j 2^j, \qquad q = \sum_{k=0}^{n} q_k 2^k \tag{1}$$

where $p_j$ and $q_k$ are qubits that can represent 0 or 1. Each bit of product $N$ is determined by summing the partial products at the corresponding position and incorporating associated carry values. The sum of partial products at bit position $i$, denoted $S_i$, is given by:

$$S_i = \begin{cases} \sum_{j=0}^{i} p_j q_{i-j}, & 0 \leq i \leq n \\ \sum_{j=i-n}^{n} p_j q_{i-j}, & n < i \leq 2n \end{cases} \tag{2}$$

The output bit $r_i$ of the product $N$ at position $i$ can be recursively computed for ascending order as follows:

$$r_i = S_i + C_{i-1} - 2C_i \tag{3}$$

where $C_{i-1}$ is the carry propagated from the previous bit position, and $C_i$ is the carry generated at the current position. The carry value $C_i$ is defined as the integer quotient obtained by the sum $S_i + C_{i-1}$ by 2:

$$C_i = \left\lfloor \frac{S_i + C_{i-1}}{2} \right\rfloor \tag{4}$$

where $\lfloor x \rfloor$ is the largest integer not exceeding $x$. This process can be effectively visualized using a binary multiplication table, where each row represents a partial product generated by multiplying $p$ with a bit of $q$ and shifting accordingly. Each vertical column contributes to $S_i$, from which both $r_i$ and $C_i$ are derived.

| P | | | | | | $p_n$ | $p_{n-1}$ | ... | $p_1$ | $p_0$ |
|---|---|---|---|---|---|---|---|---|---|---|
| q | | | | | | $q_n$ | $q_{n-1}$ | ... | $q_1$ | $q_0$ |
| Multiplication table | | | | | | $p_n$ | $p_{n-1}q_0$ | ... | $p_1q_0$ | $p_0q_0$ |
| | | | | | $p_nq_1$ | $p_{n-1}q_1$ | ... | $p_1q_1$ | $p_0q_1$ | |
| | | | | ⋰ | ⋰ | ⋰ | ⋰ | ⋰ | | |
| | | | $p_nq_{n-1}$ | $p_{n-1}q_{n-1}$ | ... | $p_1q_{n-1}$ | $p_0q_{n-1}$ | | | |
| | | $p_nq_n$ | $p_{n-1}q_n$ | ... | $p_1q_n$ | $p_0q_n$ | | | | |
| Downward carry | | $-2C_{2n}$ | $-2C_{2n-1}$ | ... | $-2C_{n+1}$ | $-2C_n$ | $-2C_{n-1}$ | ... | $-2C_1$ | $-2C_0$ |
| Upward carry | $C_{2n}$ | $C_{2n-1}$ | ... | $C_{n+1}$ | $C_n$ | $C_{n-1}$ | ... | $C_1$ | $C_0$ | |
| Result bits | $r_{2n+1}$ | $r_{2n}$ | $r_{2n-1}$ | ... | $r_{n+1}$ | $r_n$ | $r_{n-1}$ | ... | $r_1$ | $r_0$ |

Table 1. Modified binary multiplication table.

To analyze the maximum carry values, we consider the case where both $p$ and $q$ consist entirely of ones, resulting in the maximum number of ones at each bit position. This provides critical insight into the structural limits of the carry values.

As an illustrative case, consider $n = 4$, with '$p = q = 1111_2$. The corresponding binary multiplication table and associated values are:

| p | | | | | 1 | 1 | 1 | 1 |
|---|---|---|---|---|---|---|---|---|
| q | | | | | 1 | 1 | 1 | 1 |
| Multiplication table | | | | | 1 | 1 | 1 | 1 |
| | | | | 1 | 1 | 1 | 1 | |
| | | | 1 | 1 | 1 | 1 | | |
| | | 1 | 1 | 1 | 1 | | | |
| $S_i$ | | 1 | 2 | 3 | 4 | 3 | 2 | 1 |
| Downward carry | 0 | -2 | -4 | -6 | -6 | -4 | -2 | 0 |
| Upward carry | 1 | 2 | 3 | 3 | 2 | 1 | 0 | |
| Result bits | 1 | 1 | 1 | 0 | 0 | 0 | 0 | 1 |

The sequence $S_i$ exhibits a distinct symmetrical pattern defined by:

$$S_i = \begin{cases} i + 1, & \text{if } i \leq n \\ 2n + 1 - i, & \text{if } i > n \end{cases} \quad (5)$$

This characteristic of $S_i$ also influences the behavior of the carry values. In particular, both the upward and downward carry values exhibit similar symmetry due to the structure of binary multiplication and the accumulation of partial products. The carry at each bit position is calculated recursively as:

$$C_i = \left\lfloor \frac{S_i + C_{i-1}}{2} \right\rfloor \quad (6)$$

For $0 < i \leq n$, assuming the maximal values where $S_i = i + 1$ and $C_{i-1} = i - 1$, we obtain:

$$C_i = \left\lfloor \frac{(i+1) + (i-1)}{2} \right\rfloor = i \quad (7)$$

At position $i = n + 1$, where $C_{i-1} = i - 1$, the carry becomes:
$$C_i = \left\lfloor \frac{(2n + 1 - i) + (i - 1)}{2} \right\rfloor = n = 2n + 1 - i \tag{8}$$
For position $i > n + 1$, we observe the pattern $C_{i-1} = 2n + 2 - i$, which leads to:
$$C_i = \left\lfloor \frac{(2n + 1 - i) + (2n + 2 - i)}{2} \right\rfloor = 2n + 1 - i \tag{9}$$
In summary, the carry values in the all-1 case follow a piecewise function by mathematical induction:
$$C_i = \begin{cases} 0, & i = 0 \\ i, & 0 < i \leq n \\ 2n + 1 - i, & i > n \end{cases} \tag{10}$$
This structural analysis of binary multiplication and carry propagation reveals a symmetrical pattern in both partial sums and carries. As a result, we can now represent the unknown carry values as binary variables, facilitating constraint-based modeling of the factorization process.

Furthermore, understanding the upper bounds of $S_i$ and $C_i$ at each bit position allows us to introduce optimization techniques to prune infeasible solutions during the search process, significantly improving computational efficiency

## HUBO Model

To solve the integer factorization problem using quantum annealing, we formulate it as a high-order unconstrained binary optimization (HUBO) model. The approach encodes the binary multiplication of two unknown primes, $p$ and $q$ into a set of constraints over binary variables.

From each bit position $i$, the constraint ensuring correct binary multiplication is defined as:
$$S_i + C_{i-1} - 2C_i - r_i = 0 \tag{11}$$
Squaring this equation embeds the constraint as a penalty term:
$$f(S_i, C_{i-1}, C_i r_i) = (S_i + C_{i-1} - 2C_i - r_i)^2 \tag{12}$$
Where $S_i$ represents the sum of product terms contributing to bit position $i$. For base case for $i = 0$, is simplified as:
$$(p_0 q_0 - 2C_0 - r_0)^2 = (1 - r_0)^2 \tag{13}$$
since $p_0 = q_0 = 1$ for odd primes. For the general case $1 \leq i \leq 2n$, the full cost function accumulates all constrains:
$$\sum_{i=1}^{2n} (S_i + C_{i-1} - 2C_i - r_i)^2 \tag{14}$$
and the most significant carry $C_{2n}$ contributes:
$$(C_{2n} - r_{2n+1})^2 \tag{15}$$
The HUBO model for prime factorization is as follows:
$$(p_0 q_0 - 2C_0 - r_0)^2 + \sum_{i=1}^{2n} \left( \sum_{j+k=i} p_j q_k + C_{i-1} - 2C_i - r_i \right)^2 + (C_{2n} - r_{2n+1})^2 - \sum_{i=0}^{2n+1} r_i^2 \tag{16}$$

## QUBO model with constraints

To leverage the compatibility of standard quantum solvers with quadratic formulations, selected components of the HUBO model are reformulated into a quadratic unconstrained binary optimization (QUBO) model. In this reformulation, the objective function $f$ is defined at bit position $i = 1$, where higher-order terms are avoidable and the constraint structure simplifies due to known properties of the primes involved.

Since the least significant bits of two odd prime numbers $p$ and $q$ are both 1 (i.e., $p_0 = q_0 = 1$), the multiplication constraint for bit position $i = 1$ reduces to a quadratic form:
$$f = (p_1 + q_1 - 2C_1 - r_1)^2 \tag{17}$$
$$= p_1 + q_1 + 4C_1 + r_1 + 2(p_1 q_1 - 2p_1 C_1 - p_1 r_1 - 2q_1 C_1 - q_1 r_1 + 2C_1 r_1) \tag{18}$$

Since all resulting terms are either linear or quadratic, this expression can be encoded directly in a QUBO matrix. For other positions $i \neq 1$, the binary multiplication constraint naturally involves higher-order terms. To approximate these relationships while preserving solvability, we impose them as quadratic constraints in the form:

$$r_i = S_i + C_{i-1} - 2C_i \quad (19)$$

To tolerate small numerical deviations inherent in quantum annealing, this equality is relaxed into a pair of inequalities using a small tolerance $\varepsilon$:

$$k_i - \varepsilon \leq S_i + C_{i-1} - 2C_i \leq k_i + \varepsilon \quad (20)$$

Here, $k_i$ represents the known value of the $i$-th bit of the target composite number $N$. These constraints ensure local consistency with binary arithmetic, including the effects of carry propagation.

To ensure the overall accuracy of the factorization result, we impose an additional constraint on the entire binary product of two unknown primes, which we define as a global constraint. Specifically, we require that their product closely approximates the target semiprime $N$ within a small tolerance $\varepsilon$, as follows:

$$N - \varepsilon < \left(\sum_{j=0}^{n} p_j 2^j\right) \times \left(\sum_{k=0}^{n} q_k 2^k\right) < N + \varepsilon \quad (21)$$

This global constraint ensures that the full bitwise multiplication of the inferred prime candidates accurately reconstructs the composite number $N$, thereby validating the correctness of the factorization result.

# Result

We evaluated the performance of the proposed HUBO and CQM-based factorization models using quantum annealing hardware. A set of semiprime instances with increasing bit-lengths were tested, focusing on the scalability, convergence, and constraint efficacy of each formulation.

## HUBO-Based Factorization

The HUBO model of Eq. 16 encodes the full logic of binary multiplication and carry propagation into a high-order polynomial. To verify the algorism on small instances, we used the ExactPolySolver from the D-Wave Ocean SDK—a classical solver that evaluates all binary configurations in memory. This allowed exact solution verification without requiring quantum hardware (QPU).

For instance, the HUBO model successfully factored the 10-bit semiprime $N = 29 \times 31 = 899$, using approximately 16 binary variables. However, the number of binary variables increases rapidly with the bit-length $n$ of the prime factors. Specifically, the model required 4 variables for $n = 2$, 10 for $n = 3$, 16 for $n = 4$, 24 for $n = 5$, and 32 for $n = 6$. Since the exhaustive solver must evaluate all $2^n$ binary states and store them in memory, the required memory grows exponentially as the number of variables increases.

At 32 binary variables—corresponding to semiprimes derived from 6-bit factors—the number of configurations reaches $2^{32} \approx 4.3 \times 10^9$, requiring approximately 128 GB of memory to store all possible binary vectors. In our experiments, this led to a MemoryError, illustrating the practical limitations of classical brute-force solvers for HUBO models with more than 6-bit inputs. Figure 1 visualizes this exponential memory growth as a function of the bit-length of $N = p \times q$. The blue line shows the estimated memory usage based on $2^v \times v$, where $v$ is the number of binary variables. The orange bars show the corresponding number of binary variables. The red dashed line indicates the practical limit observed due to memory exhaustion.

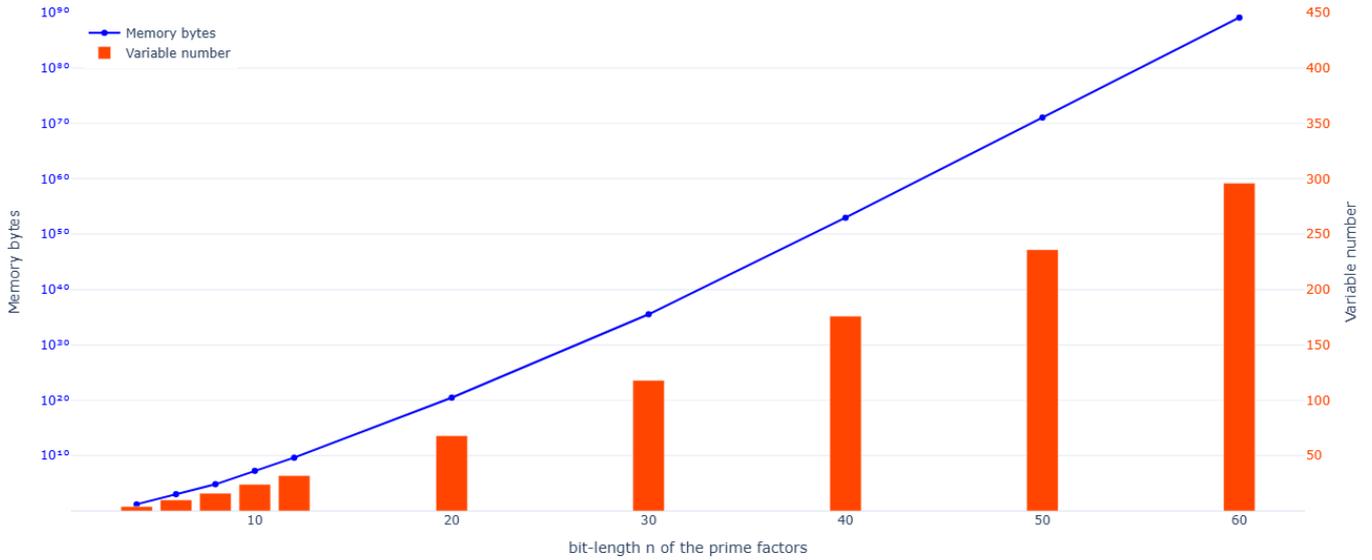

**Figure 1.** Estimated memory usage (blue line, log scale) and number of binary variables (orange bars) required by the HUBO model as a function of the bit-length $n$ of the prime factors. Memory usage is estimated as $2^v \times v$, where $v$ is the number of binary variables. This visualization illustrates the exponential growth in memory requirements and the rapid increase in model complexity, underscoring the scalability limitations of brute-force solvers for HUBO.

These results demonstrate that, although the HUBO model preserves the exact structural logic of binary multiplication, the number of required binary variables increases rapidly due to the inherent structure of the multiplication algorithm. Consequently, memory requirements scale exponentially, leading to a fundamental bottleneck in classical computation. This exponential growth indicates that the current algorithm is not well-suited for the HUBO framework at larger bit-lengths. To overcome this limitation and support practical application to larger semiprimes, we transitioned to the CQM formulation, which eliminates higher-order terms and is better aligned with quantum annealing hardware due to its reduced structural complexity.

## CQM-based factorization

While the HUBO model suffers from memory limitations when executed on a classical solver, the CQM formulation can be executed on a QPU and does not face the same restrictions. As a result, whereas HUBO was only able to factor 10-bit semiprimes due to memory limits, the CQM model using Eqs. 19, 20, and 21 as constraints and the QUBO model of Eq. 18 is successfully factored semiprimes as large as 60 bits. Specifically, we were able to factor 1,152,921,423,002,469,787 (= 1,073,741,789 × 1,073,741,783).

To evaluate scalability, semiprime instances were selected at 10-bit intervals, and each was tested with 10 independent annealing runs. The resulting number of successful factorizations was measured based on the validated condition that $p \times q = N$, regardless of intermediate carry accuracy (see Fig. 2). These results indicate that although the success rate gradually decreases with increasing bit-length, the CQM model is still capable of producing correct factorizations for considerably large semiprimes. Furthermore, as D-Wave hardware continues to improve, it is likely that even larger instances can be successfully factored with higher accuracy in the future.

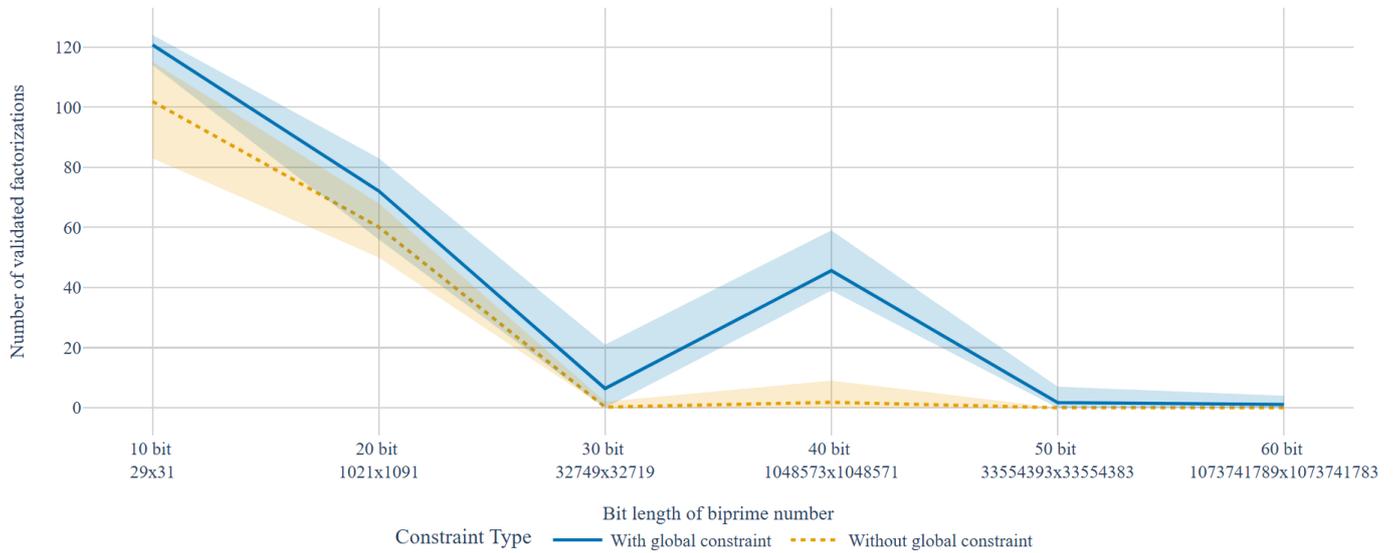

**Figure 2.** Average and range of validated factorization counts across semiprimes of 10 to 60 bits. The solid blue line represents results of new algorithm with the global constraint, which enforces an additional product-level condition. The dotted orange line corresponds to the algorithm without the global constraint, which includes only bitwise multiplication constraints. Shaded areas indicate the observed minimum–maximum values across 10 runs per setting.

## Discussion

The HUBO formulation accurately captures the logical structure of binary multiplication but exhibits poor scalability due to the combinatorial explosion of high-order terms with increasing bit-length. In practice, we found that the HUBO model becomes impractical for semiprimes exceeding approximately 10 bits, as the generated polynomial quickly surpasses the limitations of current quantum annealing hardware in terms of both runtime and qubit connectivity. To address these scalability constraints, we transitioned to the CQM, explicitly separating the optimization into an objective function and associated constraints. This restructuring substantially reduces model complexity, ensuring compatibility with existing quantum annealers. Adopting the CQM framework thus represents a practical compromise between exact logical fidelity and computational efficiency, enabling successful factorization of significantly larger semiprimes.

The beneficial impact of constraint inclusion on solution quality is clearly evident in Figure 2, indicating that the global constraint effectively guide the annealing process toward more accurate solutions by limiting the search space to regions near valid results. This observation is further supported by analyzing the factorization error—defined as the absolute difference between the computed product ($p \times q$) and the actual target value N—which consistently decreases with the application of the global constraint. These results underscore the critical role constraints play in enhancing the efficacy of the search process.

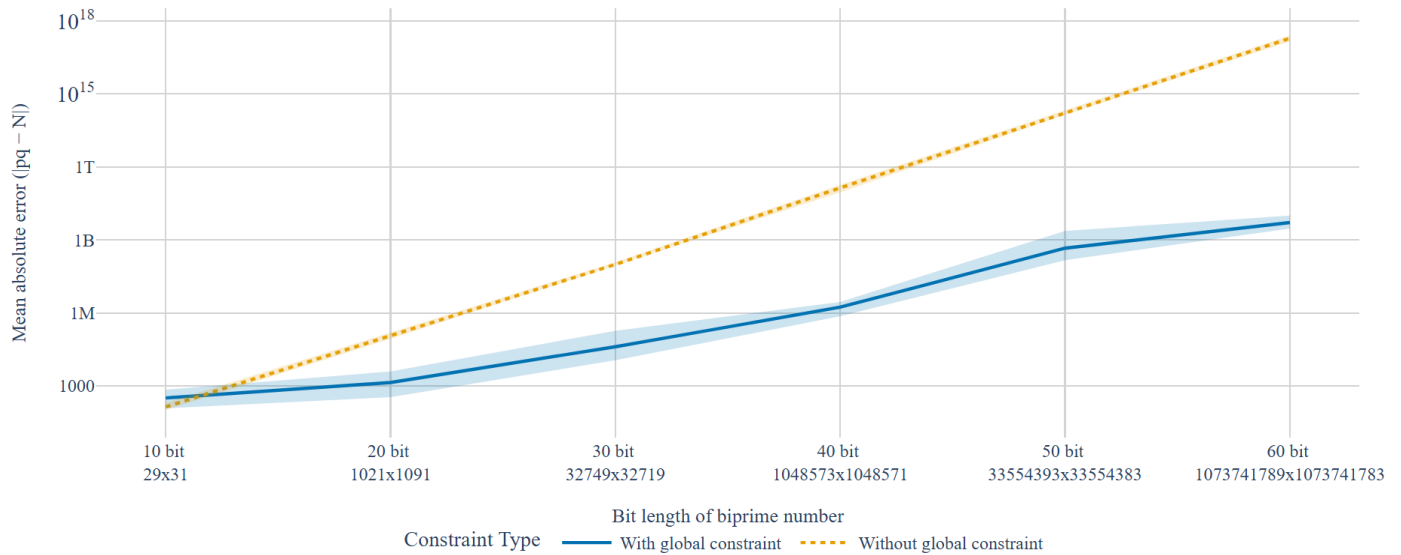

**Figure 3.** Mean absolute error ($|pq - N|$) across semiprimes of increasing bit-lengths, plotted on a logarithmic scale. The solid blue line represents results our algorithm with the global constraint, while the dotted orange line shows results without the global constraint. Shaded regions indicate the minimum–maximum error range observed over 10 independent runs per bit-length. The logarithmic y-axis emphasizes the difference in error magnitude, particularly for larger problem sizes.

To quantitatively evaluate the impact of constraints, we analyzed the magnitude and distribution of factorization errors across various semiprime bit-lengths, comparing scenarios with and without the global constraint. Figure 3 presents the mean absolute error ($|p \times q - N|$) on a logarithmic scale, clearly demonstrating that errors are significantly reduced when the global constraint are applied. The shaded areas illustrate the observed error range across multiple independent trials, emphasizing the consistency of this improvement. The error trends highlighted in Figure 3 confirm that incorporating the global product constraints systematically reduces errors for all tested bit-lengths. These findings indicate that constraint application not only improves solution accuracy but also enhances the precision of quantum annealing by restricting the solution space to feasible regions. Overall, our results highlight the importance of strategic constraint formulation in quantum optimization and validate CQM as a robust and scalable method for quantum-based integer factorization.

### ACKNOWLEDGEMENTS
D-Wave leap quantum cloud service was supported by the Chungbuk Quantum Research Center at Chungbuk National University. A. R., K. K., and K. J. was supported by the MSIT(Ministry of Science and ICT), Korea, under the ITRC(Information Technology Research Center) support program(IITP-RS-2024-00437284) supervised by the IITP(Institute for Information & Communications Technology Planning & Evaluation. A. R. and K. K. was supported by Creation of the quantum information science R&D ecosystem(based on human resources) (Agreement Number) through the National Research Foundation of Korea(NRF) funded by the Korean government (Ministry of Science and ICT(MSIT)) (RS-2023-00256050).

### DATA AVAILABILITY
The Python code used in this paper is in the supplementary material.

Supplementary 1. Python code for HUBO model.

Supplementary 2. Python code for QUBO model with constraints.

### COMPETING FINANCIAL INTERESTS
The authors declare no competing interests.